# ConceptModeller: a Problem-Oriented Visual SDK for Globally Distributed Enterprise Systems

Sergey V. Zykov
ITERA Oil and Gas Company
Moscow, Russia
E-mail: szykov@itera.ru

**Abstract**[1]

The paper describes problem-oriented approach to software development. The approach is a part of the author's integrated methodology of enterprise Internet-based software design and implementation. All aspects of software development, from theory to implementation, are covered.

## 1. Introduction

The present-day society has accumulated a huge and ever-growing data bulk, manipulating which is quite an issue, particularly, because of its heterogeneous and weak-structured character. Under socio-economic globalization, efficient information system (IS) functioning requires development of novel design and implementation concepts and methodologies.

Accelerated IT penetration into each and every sphere of human activity demands heterogeneous IS data and metadata (MD) integration, based on often contradicting concepts, methodologies, models and approaches (including those created by major SDK producers, such as Microsoft, IBM, Oracle, SAP, BEA, etc.). A number of leading research teams are approaching the unification problem of theories, languages and tools for large-scale IS design, implementation and maintenance. However, a satisfactory solution has not been obtained yet.

Thus, conceptually and methodologically comprehensive formal models and SDK should be created to integrate efficient IS maintenance throughout the entire lifecycle.

Therewith, a set of interrelated problems arises, including application development methodological diversity and inadequacy to standards, incompatible mathematical foundations, (meta)data object definition languages, and unbalanced user-to-IS and user-to-DBMS interfaces.

Research objective is analysis, reasoning and development of conceptual and methodological foundations of full-scale IS construction (prototyping, implementation and maintenance) and its application to enterprise information collection, processing and reporting within global network environment.

Major research tasks include analysis, systematization and generation of a conceptual approach for continuous integrated design, implementation and support for globally distributed IS, including the supporting methodology, (M)DO formal model for problem domain and computational environment, and problem-oriented CASE-and-RAD-tools, interface and architecture solutions for prototypes and full-scale portals.

## 2. Theoretical Background

Research methods are based on creative synthesis of fundamental concepts of finite sequence [1], category [4], computations and semantic network [5] theories.

The paper claims to be the first to reason, present and practically approve conceptual and methodological aspects of continuous integrated enterprise IS development for global network environments that cover the entire lifecycle.

All the approaches known as yet either have methodological "gaps" or do not result in satisfactory enterprise applications meeting the major criteria levels (scalability, extendibility, fault tolerance, etc.).

As a result of analytical problem domain research, the author has created an innovative comprehensive *conceptual* approach to IS integrated design and implementation, which is a solution of a major scientific problem of primary economic importance and which comprises the following elements:

1. Formal model set (conceptual problem domain model [11], abstract machine model for SDK and computational environment [7]);

---

2. IS design, implementation and support methodology [8];
3. SDK choice criteria set for application prototyping, design and implementation [11];
4. New SDK (*ConceptModeller* for visual problem-oriented design, content management IS) [7,8].

*Conceptual* IS design, one of essential components of the approach, unifies (M)D computational model in terms of (meta)data objects ((M)DO), language and application tools for manipulating them. Another integral part of the approach is the development methodology that supports "non-stop" multi-level two-way iterative software design and implementation and controls (M)DO actuality, completeness, consistency and integrity throughout the entire IS lifecycle. One more major research aspect is related to development applications for CASE, RAD and integration of (M)DB and IS.

Under *primary concepts* physical or abstract objects selected within the problem domain are implied. Primary concepts are intensional objects, i.e. types or sorts interpreted as sets.

Under *constants* (or instantiations) separate individuals of interpreted primary concepts are implied. Generally speaking, a constant is an intensional object, however under a certain instantiation it should be bound to a single interpretation of a corresponding type.

Under an *interpretation* $|D_i|$ of primary concept $D_i$, all constants from $D_i$ are bound to respective instantiations from $|D_i|$.

Any (free or bound) variable should have an assigned type (for free variables this is marked by a *type arc*, i.e., a "t"-marked arc pointing to a primary concept, which corresponds to the type(s) of constant(s) that can be assigned to the variables).

Under a *frame* a graph is implied, which represents a knowledge unit in terms of access and processing.

Under a *simple frame* a frame is implied, which contains no sub-graphs, but only constants, variables and arcs. Frames correspond to relations at DB level.

Arc type and interpretation frames are structurally subdivided into event, function (predicable) and characteristic ones.

*Event frames* are action models. *Events* are treated as specific predicates. A network representation of an elementary event frame is a node corresponding to the event predicate and outgoing role-marked *role arcs* pointing to the nodes representing predicate arguments. The arguments are either typed variables or typed constants of primary concept(s). Event frames are templates (i.e. intensional objects), since their arguments are not evaluated and, consequently, do not carry information on the real event. However, upon evaluation of all variables by appropriately typed constants, the predicate (and event frame) gets a Boolean value. Role arcs denote event arguments and imply argument meaning in the event. This is the major distinction between role arcs and logical predicate arguments.

Event frame role denotations are described in Table 1.

**Table 1. Event frame roles**

| Short denotation | Long denotation | Semantics interpretation |
|---|---|---|
| a | agent | Action initiator (actor) |
| o | object | Action addressee |
| s | source | Action addressee location before the event |
| d | destination | Action addressee location after the event |
| r | result | Action result |

## 3. ConceptModeller: a Challenging Tool for the Integrated Methodology

The development methodology transforms IS specifications from problem domain concepts to formal model entities, further, by CASE tools, to frame system and OR(M)DB scheme, and finally, to formal architecture-and-interface description of target IS. The methodology includes semantically driven algorithm of iterative IS component integration with reengineering capability.

Therewith, the "non-stop", continuous character of the methodology is provided due to the innovative *ConceptModeller* CASE tool for problem domain conceptual design, developed under author's supervision in Moscow Engineering Physics Institute (MEPhI).

*ConceptModeller* CASE tool is aimed at automated translation of the problem domain (M)DO model to UML specification with its later conversion to target (M)DB and IS schemes.

Major *ConceptModeller* features are:

- Formal model adequacy to problem domain;
- Problem orientation (user operates in natural language object-relational terms);
- Visibility (visual development is used);
- State-of-the-art IS development methodologies support (UML, BPR);
- Connectors to officially recognized and industrially approved CASE and RAD tools (*IBM Rational, Microsoft Visual Studio,* etc.);
- Two-way IS development.

Due to the above benefits, *ConceptModeller* supports wide problem domain spectrum and allows modeling in nearly natural language terms (see examples below).

Moreover, *ConceptModeller* allows to automatically translate (M)DB and IS schemes into the related conceptual model during the entire lifecycle (problem domain modeling, CASE and RAD development, testing, maintenance and customization). As a result, a possibility arises to test and verify IS by pure mathematical or programming methods (e.g., an SML[2] language program) at arbitrary abstraction level.

## 4. ConceptModeller: Design and Implementation Features

*ConceptModeller* interface for simple frame design is presented in figure1.

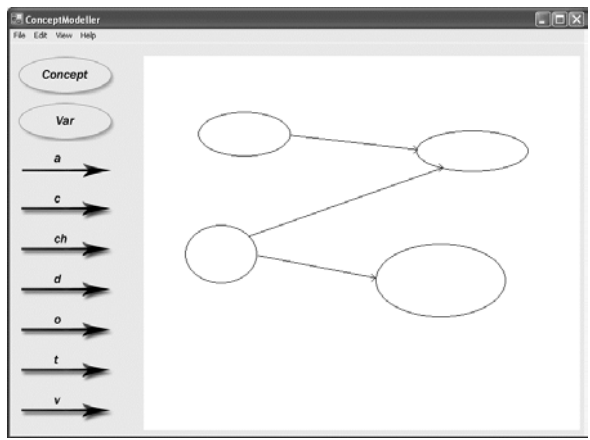

**Figure 1.** *ConceptModeller* user interface

Frame visualization interface is intuitively close to major vector graphics design software, similar to Adobe PhotoShop. Particularly, simple frame visualization toolkit includes various types of concepts and arcs (e.g., "t"-arcs connect variables and types).

Double representation of frames in graphical and DB structure form requires frame storage format development, which possesses properties of completeness, extendibility for adding metadata, and unique visual interpretation (including multiple instance case).

Since Microsoft .NET has been chosen as the development platform, XML is a reasonable DB format solution. XML data structures provide convenient visualization and frame-to-UML translation.

Frame elements have the following parameters: identifier, type, name, coordinates, hierarchy links to predecessor and successor, and a number of optional fields.

---

[2] A free SML.NET functional program language compiler (the language originated as a proof-building tool) is implemented by an Edinburgh University team and has been used by the author in MEPhI Computer Science course, supported by Microsoft Research Ltd. in 2002-2003 [16].

The XML DB manipulation is based on *XML Designer* component embedded into *Microsoft Visual Studio 2005* and used for frame template DB generation by the XML schema. XML file contains complete DB description, a frame element representation example is presented in figure 2.

```
<?xml version="1.0" standalone="yes" ?>
- <NewDataSet>
  - <Elements>
      <Id>1</Id>
      <Type>Var</Type>
      <Name>MyVar</Name>
      <Left>100</Left>
      <Top>100</Top>
      <Width>100</Width>
      <Height>50</Height>
      <Prev>0</Prev>
      <Next>0</Next>
      <Description>No Description</Description>
    </Elements>
  </NewDataSet>
```

**Figure 2.** XML description of a frame fragment in visualization DB

The above XML code describes variable *MyVar* of type *Var* and visual size of 100 by 50 pixels.

A simple frame visualization example by *ConceptModeller* is given in figure 3. It visualizes a frame of "supply" event of a CANDIDATE by personnel MANAGER to EMPLOYER.

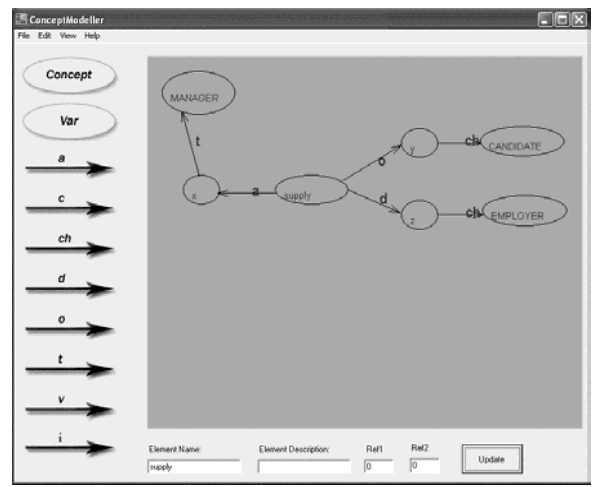

**Figure 3.** Simple frame visualization example by *ConceptModeller*

*ConceptModeller* tool is implemented in C# language and *Microsoft Visual Studio 2005* environment and contains over 4500 lines of source code including event-driven components of frame visualization, their translation to UML code and visualization of the resulting code.

Particularly, visualization component contains frame element behavior and visualization methods code. All the GUI elements are created using *Adobe Photoshop* CS editor and are imported as resources.

## 5. Methodology Application Overview

The presented methodology has been practically approved in ITERA International Group of Companies. Application architecture supports integrated (meta)data storage. The methodology transforms problem domain frame model generated by *ConceptModeller* to UML diagrams, further, by CASE-tools, to ERD (or by an abstract content management machine – into its code), and finally into target IS and (M)DB attributes.

Architecture, interface, fast prototype and full-scale application for HR and portal-based information resource management have been implemented on the basis of the conceptual approach and the methodology suggested.

Practical value of the results obtained is derived from large IS development benefits by the methods introduced. Implementation time is substantially reduced in comparison with commercial software available. Essential cost reduction for data maintenance, fault tolerance and integrity support takes place, IS customization and performance optimization have also become much easier.

The author's concepts and methodology have been used as a foundation for a number of enterprise applications development in ITERA Group with nearly 10,000 employees: *UniQue* HR ERP (1998-2000), content management IS (2001-2002), official Internet site (www.itera.ru, 2003-2004) and enterprise Intranet portal (2004). IS management, by expert estimates, annually saves several hundreds of thousands of US dollars and increases (meta)information handling efficiency.

## 6. Implementation Results Summary, Recommendations and Prospects

As a result of implementation based on formal models and methodology suggested, implementation terms and costs compared to commercial software available have been considerably reduced and functional features have been extended.

Practical implementation experience proved actuality, innovation and efficiency of the approach on the whole and separate concepts, models, tools and applications developed.

Exhaustive study of the models supporting enterprise portals is supposed to take a considerable extra time and labor resources.

The author is going to continue his studies of the formal models that support enterprise portals and their practical application.